\documentclass[useAMS]{mn2e}

\title[Latitude migration of solar filaments] {Latitude migration of solar filaments}
\author[K. J. Li]
{K. J. Li$^{1,2}$\thanks{E-mail: lkj@mail.ynao.ac.cn} \\
$^{1}$National Astronomical Observatories/Yunnan Observatory, CAS, Kunming 650011, China \\
$^{2}$Key Laboratory of Solar Activity, National Astronomical
Observatories, Chinese Academy of Sciences
  }
\begin{document}



\maketitle

\label{firstpage}

\begin{abstract}
The Carte Synoptique  catalogue of solar filaments from March 1919
to December 1989, corresponding to complete cycles 16 to 21 is
utilized to show {\bf latitudinal migration of filaments at low latitudes (
less than $50^{0}$)}, and the latitudinal
drift of solar filaments in each hemisphere in each cycle of the
time interval is compared with the corresponding drift of sunspot
groups. The latitudinal drift of filaments obviously differ from
that of sunspot groups. At the beginning of a cycle filaments
(sunspot groups)  migrate from latitudes of about $40^{\circ}$
($28^{\circ}$) with a drift velocity of about $2.4 m/s$ ($1.2 m/s$)
toward the solar equator, reach latitudes of about $25^{\circ}$
($20^{\circ}$) 4 years later at the cycle maximum with a drift
velocity of about $1.0 m/s$ ($1.0 m/s$), and halt at about
$8^{\circ}$ ($8^{\circ}$) at the end of the cycle. When
 solar activity is programming into a solar cycle, the difference
 between  the appearing latitudes of filaments and sunspot groups
becomes smaller and smaller. The difference rapidly decrease in the
first and last $\sim4$ years  of a cycle, but
almost does not decrease in the $\sim4$ years after the maximum time of
the cycle. For filaments, the latitudinal drift velocity decreases
in the first $\sim 7.5$ years of a cycle, but it increases in the
last $\sim 3.5$ years of the cycle. However, for sunspot groups the
drift velocity always decreases in a whole cycle. The different between
the latitude drift of filaments in the northern and southern
hemispheres is found not to be obvious. The latitudinal drift velocity of
filaments
 slightly differs  from each other in the northern and southern hemispheres in
a cycle. The physical implication behind the latitudinal drift of
filaments is explored.
\end{abstract}


\section{INTRODUCTION}
It is widely believed that the Sun's magnetic field is generated by
a magnetic dynamo within the Sun. Solar dynamo models  demonstrate
that the solar activity cycle displays a recycling of the two main
components of the Sun's magnetic field (Parker 1955): the poloidal
and toroidal components.
The solar meridional flow, the flow of material
in the meridional plane from the solar Equator towards the Sun's
poles at the surface and from the poles towards the Equator deep
inside the Sun to carry the dynamo wave towards the Equator,
should play an important role in the Sun's magnetic
dynamo  (Parker 1987; Choudhuri,
 Schussler,  $\&$ Dikpati 1995; Durney 1995; Hathaway et al.
2003; Charbonneau 2007).
Propagation of the toroidal field
wave at the base of the solar convection zone is believed to
manifest itself on the surface as the migration of the sunspot
activity belt toward the Equator (the so-called butterfly diagrams),
 (Hathaway et al. 2003).
It is helpful to investigate the latitude migration of sunspot
activity for understanding the Sun's magnetic dynamo. Hathaway et
al. (2003) examined the drift of the centroid of the sunspot area
toward the Equator in each hemisphere from 1874 to 2002 and found
that the drift rate slows as the centroid approaches the equator,
which was also found early by Li, Yun, $\&$ Gu (2001) and Li et al.
(2002). Hathaway et al. (2003) compared the drift rate at sunspot
cycle maximum with the period of each cycle for each hemisphere and
found a highly significant anti-correlation: hemispheres with faster
drift rates have shorter periods. These observations are consistent
with a meridional counterflow deep within the Sun as the primary
driver of the migration toward the Equator and the period associated
with the sunspot cycle (Hathaway et al. 2003).

Solar filaments have been observed since the invention of the spectrohelioscope
(Tandberg-Hanssen 1995; Heinzel 2007).
They are prominences projected against the solar disk, displaying themselves as cool
and dense ``clouds in the solar corona" (Kiepenheuer 1953;  Tandberg-hanssen 1995, 1998).
The shape, fine structure, dynamics, and physical
properties of solar filaments vary one to another in the wide range of values
(Hundhausen,  Hansen,  $\&$ Hansen 1981;
Illing $\&$ Hundhausen 1986; Heinzel 2007; Lin,  Martin, $\&$ Engvold 2008).
{\bf It is always interesting to study the general trend of the solar activity by
using statistical data such as the ``Cartes synoptiques" of solar filaments 
(d¡¯Azambuja 1923; Mouradian 1998a, 1998b).
Mein and Ribes (1990)  used the ``Cartes synoptiques" and evidenced the meridional circulation
and the existence of convective giant cells by following the filaments used
as indicators of magnetic tracers, and  
it is very important to understand how the solar dynamo works and therefore
the existence of magnetic solar cycles.
The phenomenon ``rush to the polar"
 has been confirmed through studying the migration of polar filaments towards the pole at the
beginning of solar cycles by many authors (Topka et al 1982; Makarov $\&$ Sivaraman 1989;
 Mouradian $\&$ Soru-Escaut 1994; Shimojo et al. 2006; Li et al. 2008).}
Solar filaments are distributed on the whole solar surface, from the solar equator
to the poles, and during the whole period of a sunspot cycle (Li et al 2007).
Such a distribution is not
random due to that solar filaments are closely connected with sites of magnetic fields on
the solar disk (Martin 1990).  
{\bf The north\,--\,south asymmetry of filaments in solar cycles 16\,--\,21 is investigated with the use of the
``Cartes synoptiques" (Li et al. 2010), and filament activity is found regularly
dominated in each of cycles 16\,--\,21 in the same hemisphere as that inferred by sunspot activity
(Li,  Schmieder, $\&$ Li 1998; Li et al. 2002).}

Filaments' feature to occur in all heliospheric latitudes
and to outline the boundary between magnetic fields with different polarities makes themselves
suitable tracers for the large-scale pattern of the weak background magnetic field of the Sun
(McIntosh 1972; Low 1982; Minarovjech et al. 1998a, 1998b). On the other hand, study of the
occurrence of filaments can help us to better understand distribution of these fields
on the solar disk, their development within a solar cycle, and especially, provide
useful insights into the nature of the Sun's magnetic field (Rusin et al. 1998, 2000;
Mouradian $\&$ Soru-Escaut 1994). Study on filaments (prominences) through both individual events
and statistical analyses are of importance, and some progresses on the knowledge of
filaments have been achieved up to now, including both the morphological aspects and theoretical scenarios
of filaments (Anzer 2002; Engvold 2004;  Heinzel 2007; Schmieder et al. 2007, 2009; 
{\bf Labrosse et al. 2010; Mackay et al. 2010}; and references therein).
In this paper, we
will investigate the latitude migration of solar filaments, focusing on their statistical properties,  and further compare with that of sunspots.

\section{OBSERVATIONS, ANALYSES, AND RESULTS}
The different solar activity time series analyzed in our study
are:

\,--\,The first one: the Carte Synoptique solar filaments archive,
namely the catalogue of solar filaments, observed at Meudon from March 1919 to December
1989, corresponding to Carrington solar rotation numbers 876 to
1823, which can be accessed via the NOAA's  web
site\footnote{ftp://ftp.ngdc.noaa.gov/STP/$SOLAR_{-}DATA\\/SOLAR_{-}FILAMENTS$}.
The observations of filaments were mainly shown in maps drawn at Meudon, which were a synthetic representation of
solar active regions and filaments (d¡¯Azambuja 1923), and the Meudon maps were complemented with tables of solar
active regions and filaments (for datails, see Mouradian 1998a, 1998b).  The World Data
Center A (WDC-A) for Solar-Terrestrial Physics has digitized the Carte Synoptiques
(Coffey $\&$ Hanchett 1998).

\,--\,The second: the observational data of sunspot groups, which
come from the augmented Royal Greenwich Observatory (RGO) data set
(the RGO data set extends from 1874 to 1976; thereafter, the
observations are from NOAA) and are available at NASA's Web
site\footnote{http://solarscience.msfc.nasa.gov/greenwch.shtml}. The
data set comprises sunspot groups during the period of May 1874 to
March 2009 and will be updated monthly.

The normal solar activity is usually applied to solar active events
whose latitudes are lower than $50^{\circ}$ (Sakurai 1998; Li et al.
2008), thus used here are filaments whose latitudes are lower than
$50^{\circ}$. Fig.  1 shows both filaments' and sunspot groups'
positions versus time, namely the familiar butterfly diagrams. The
figure indicates how the bands of filaments (sunspot groups) drift
toward the equator and how successive cycles overlap during the
minimum time of a sunspot cycle. As the figure displays, filaments
are distributed themselves at higher latitudes and within a wider
latitude band at a certain time epoch than sunspot groups do.  It is
difficult to accurately divide sunspot groups into solar cycles to
which they really belong (Harvey 1992)  and  even more difficult to
divide filaments into solar cycles. A criterion for how a solar
activity event can be assigned to an active cycle was once proposed
by Li, Yun, $\&$ Gu (2001). Here, according to the criterion and
with the use of the Carte Synoptique solar filament archive,
filaments are divided into individual butterflies. Then monthly mean
latitudes of filaments in the northern and southern hemispheres are
calculated and plotted in Figs. 2 and 3, respectively. For sunspot
activity (sunspot groups or sunspot areas), a second-order
polynomial curve could give a satisfactory fit to the monthly mean latitudes
of sunspot activity (Li Yun, $\&$ Gu 2001; Hathaway et al. 2003),
but for filament activity, it is a third-order not second-order
polynomial curve that can give a satisfactory fit to the monthly mean
latitudes of filaments respectively in the northern and southern
hemispheres, and the fitting curves are also given in Figs. 2 and
3. In order to compare the latitude drift of filaments with that of
sunspot groups, the best fit to the monthly mean latitudes of
sunspot groups respectively in the northern and southern hemispheres
in each of cycles 16 to 21 is given here in Figs. 2 and 3, respectively.  As
the two figures show, the latitudinal drift of filaments obviously
differ from that of sunspot groups: the monthly mean latitude of
filaments is higher than the corresponding latitude of sunspot
groups in both the northern and southern hemispheres at each month
of a cycle, but the difference between the two  becomes smaller and
smaller with solar activity of a cycle  programming into the cycle.

In order to compare the latitude drift of filaments respectively in
the northern and southern hemispheres, the unsigned monthly mean
latitude of filaments is plotted in Fig.  4. As the figure shows,
the different between the two is not obvious in a cycle for cycles
16 to 21 except cycle 19.

The unsigned monthly mean latitudes of filaments
for all of the 6 solar cycles are plotted together in a cycle in Fig.  5 by using the time
measured relative to the maximum  times of their corresponding cycles. Also given in the figure is a third-order polynomial
fit to these points. As the figure shows, in a statistical view filaments  appear
at latitudes of about $40^{\circ}$ at the beginning of a cycle, move
toward the equator,  reaching $25^{\circ}$ some 4 years later at the
cycle maximum, and then halting at about $8^{\circ}$ at the end of the
cycle.

Similarly, the unsigned monthly mean latitudes of sunspot groups in the 6 solar cycles are plotted
 together in a cycle in Fig.  5
  using the time measured relative to the  maximum times of their corresponding
cycles.  A second-order polynomial is used to fit  these points,
which is also given in the figure. Then we know the difference between the
two fits varying with the time measured relative to cycle maximum
time, which is shown in the bottom panel of the figure as well. Filaments appear at much higher
latitudes than sunspot groups do at the beginning of a cycle; when
 solar activity is programming into the cycle, the difference
 between  the appearing latitudes of filaments and sunspot groups
becomes smaller and smaller. The difference rapidly decrease in both the
first and last $\sim4$ years  of a cycle, but
hardly  decrease in the $\sim4$ years after the maximum time of the
cycle.

Shown in Figs. 6 and 7 are the latitudinal drift velocity of filaments varying
with time in a cycle in the northern and southern hemispheres, respectively. The latitudinal drift
velocity ranges within about 1.5 to  3.5m/s in a cycle, and it
slightly differs  from each other in the northern and southern
hemispheres in a cycle.
Also displayed in the two figures is the latitudinal drift velocity of sunspot groups
varying with time  in each of cycles 16 to 21 and in the northern and southern hemispheres, respectively.
As the two figures
show, the drift velocity linearly decreases with time for sunspot
groups in both the northern and southern hemispheres in each of
cycles 16 to 21, but for filaments it firstly decreases  and then
increases in a cycle. Hathaway et al. (2003) investigated the drift
of the centroid of the sunspot area toward the equator in each of
the northern and southern hemisphere from 1874 to 2002 and found
that the drift rate slows as the centroid approaches the equator
except cycle 21 in the northern hemisphere. Fig.   8 shows the
general feature of the latitudinal drift velocity in a cycle
respectively for filaments and sunspot groups when the unsigned
monthly mean latitude of filaments and sunspot groups for each
hemisphere and for each cycle is plotted together by
using the time measured relative to the maximum time of that cycle
(see Fig.   5). At the beginning  of a cycle filaments (sunspot
groups) migrate from latitudes of about $40^{\circ}$ ($28^{\circ}$)
with a drift velocity of about $2.4 m/s$ ($1.2 m/s$) toward the
solar equator, reach latitudes of about $25^{\circ}$ ($20^{\circ}$)
4 years later at the cycle maximum with a drift velocity of about
$1.0 m/s$ ($1.0 m/s$), and halt at about $8^{\circ}$ at the end of
the cycle. For filaments the latitudinal drift velocity decreases in
the first $\sim 7.5$ years of a cycle, but it surprisingly increases
in the last $\sim 3.5$ years of the cycle.

The latitudinal drift velocity varying with latitude is plotted in
Fig.   9 respectively for filaments and sunspot groups in each
hemisphere in each cycle. For filaments, the drift velocity
decreases in a cycle with latitudes until that they reach latitudes
of about $20^{\circ}$, after then it increases with latitudes until
the ending of the cycle. For sunspot groups, drift velocity
decreases with latitudes in each hemisphere in each of cycles 16 to
21 without any exception.

The latitudinal drift acceleration varying with time is plotted in
Fig.   10 respectively for filaments and sunspot groups when the
unsigned monthly mean latitude of filaments and sunspot groups for
each hemisphere and for each cycle is plotted together by using the time measured relative to the maximum time of that
cycle (see Fig.   5). For sunspot groups, the acceleration is a
constant, about -0.2 $(degrees/s^{2})$ in a cycle, but for
filaments, the acceleration linearly increase from about -1.2
$(degrees/s^{2})$ at the beginning of a cycle to about 0.4
$(degrees/s^{2})$ at the ending of the cycle.

\section{Conclusions and Discussions}
{\bf Shimojo et al (2006) 
showed well the comparison of prominence activities and the butterfly
diagram obtained from synoptic Carrington maps of Kitt Peak magnetograms.
The drift of prominences follows the polar ward motion of the magnetic flux
until the reversal of the polar polarity. This is well visible for latitudes $30^{0}$ to
$90^{0}$}. Latitudinal migration of filament activity over
the solar full disk was qualitatively introduced by Li et al (2008).
In the present study,
the catalogue of solar filaments from Carrington solar rotation
numbers 876 to 1823, corresponding to complete cycles 16 to 21 is
used to investigate latitudinal migration of {\bf filaments at low latitudes (less than $50^{0}$)}, 
and the latitudinal
drift of solar filaments in each hemisphere in each cycle of the
time interval is compared with that of sunspot groups in the same
time interval. The latitudinal drift of filaments obviously differ
from that of sunspot groups: for sunspot groups a second-order
polynomial can give a satisfactory description of the latitudinal drift in a
cycle, but for filaments it is a third-order not second-order
polynomial curve that can give a satisfactory description. At the beginning
of a cycle filaments migrate from latitudes of about $40^{\circ}$
with drift a velocity of about $2.4 m/s$ toward the solar equator,
reaching latitudes of about $25^{\circ}$ 4 year later at the cycle
maximum with a drift velocity of about $1.0 m/s$, and halting at about
$8^{\circ}$ at the end of the cycle. Compared with filaments,
sunspot groups appear at much lower latitudes at the beginning of a
cycle; when
 solar activity is programming into a cycle, the difference
 between  the appearing latitudes of filaments and sunspot groups
becomes smaller and smaller. The difference rapidly decrease in the
first and last $\sim4$ years and  of a cycle, but
almost does not decrease in the $\sim4$ years after the maximum time of
the cycle.

The reasons why filaments statistically appear at higher latitudes
than sunspot groups are suggested: (1) The decay of old sunspots
usually diffuses towards the solar poles (the so-called ``rush to
the pole"), an emergence of new sunspots at a certain latitude acts
with the decay component of old sunspots, which should form a
filament (or filaments) at a higher latitude. (2) The emergences of
the magnetic field usually form sunspots, but sometimes they form
ephemeral regions (tiny ``pores'), do not form sunspots. Ephemeral
regions statistically locate at higher latitudes than sunspots
(Harvey 1992). Filaments, relating with these ephemeral regions
should thus appear at higher latitudes. And (3) the background
poloidal magnetic field should be more easily observed at high
latitudes. The toroidal magnetic field of sunspot groups at a
latitude should act with the background poloidal magnetic field to
form filaments at a higher latitude.

For filaments, the drift velocity decreases in the first $\sim 7.5$
years of a cycle, but it surprisingly increases in the last $\sim
3.5$ years of the cycle. However, for sunspot groups the drift
velocity always decreases in a whole cycle. The increase of the
latitudinal drift velocity of filaments at low latitudes should
possibly imply the existence of  interaction between the paired
wings of a butterfly, such as cross-hemispheric coupling
(Charbonneau 2007). In fact, for filaments the paired wings of a
butterfly are hardly distinguished from each other near the equator
(see Fig.  1). Although sunspots are hardly observed near the
equator, filaments can be observed to cross the equator, known as
transequatorial filaments (Wang 2002), also supporting the
implication. The latitudinal drift acceleration of filaments changes
from the negative sign during the first about 7.5 years  of a cycle
to the positive sign during the last about 3.5 years of the cycle,
implying that the  mechanism driving filaments to drift should be
different in the two time intervals of a cycle.

The different between the latitude drift of filaments in the
northern and southern hemispheres is found not obvious in a cycle
for cycles 16 to 21 except cycle 19. The latitudinal drift velocity
of filaments
 slightly differs  from each other in the northern and southern hemispheres in
a cycle.

Filaments are large magnetic structures in the corona,
but sunspots are the emergence of strong magnetic field from
the Sun's interior to the photosphere and chromosphere.
The difference between the latitudinal drift of filaments
and sunspots implies that: (1)filaments should be related with
weak background magnetic field sometimes; (2) to what extent
does the latitudinal drift
of sunspots reflect the  meridional flow below the solar atmosphere?
it is an open issue.

In this study,
latitudinal drifts are determined from
the tabulated heliocentric positions of the centroids of
sunspots and filaments. Centroids from sunspot images are a good measure of
sunspot position because the dark spots are compact and clearly visible in
white light images. In contrast, filaments are long sinuous structures on the disk,
their heliocentric positions are thus less accurately determined than sunspots' positions.
Filaments are known to tend to be
aligned more east-west on the disk, this should reduce the contest of the determinations of their
latitudes. Average of filaments' and sunspots' latitudes over one month is used in this study,
this should further reduce the contest coming from the determinations of filaments'
latitudes.  In this study, we investigate the latitudinal drift of filaments in a whole solar cycle, not focusing on individual monthly averages of filaments' latitudes,
the effect of the contest on our findings should be reduce further again. Thus,
the influence of the contest in the determinations of filaments' positions should be very small and can be neglected.

{\bf One point should be emphasized. We concentrate on the behaviour of a particular class of filament. We
study only the active region (AR) or intermediate class filaments (see the
classification by MacKay et al. 2010). 
The fact that filament gravity
centers are located higher than the AR means that filaments are
preferentially located at the periphery of AR.
Therefore, the large
drift of filaments located at latitudes around $40^{0}$  possibly compared to the drift
of sunspots at the beginning of the solar cycle could  in fact reflects the fact
that there is still a mixture of polar crown filaments migrating towards the
pole and low latitude filaments following AR.}

\section*{Acknowledgments}
We thank the referee for his/her careful reading of the manuscript
and constructive comments, which improved the original version.
Data used here are
all downloaded from web sites. The authors would like to express
their deep thanks to the staffs of these web site.  The work is
supported by the NSFC under Grants 10583032, 10921303, and 40636031, the
National Key Research Science Foundation (2006CB806303), and the
Chinese Academy of Sciences.

\newpage

\input{epsf}
\begin{figure*}
\begin{center}
\epsfysize=12.cm\epsfxsize=12.cm \hskip -5.0 cm \vskip 30.0 mm
\epsffile{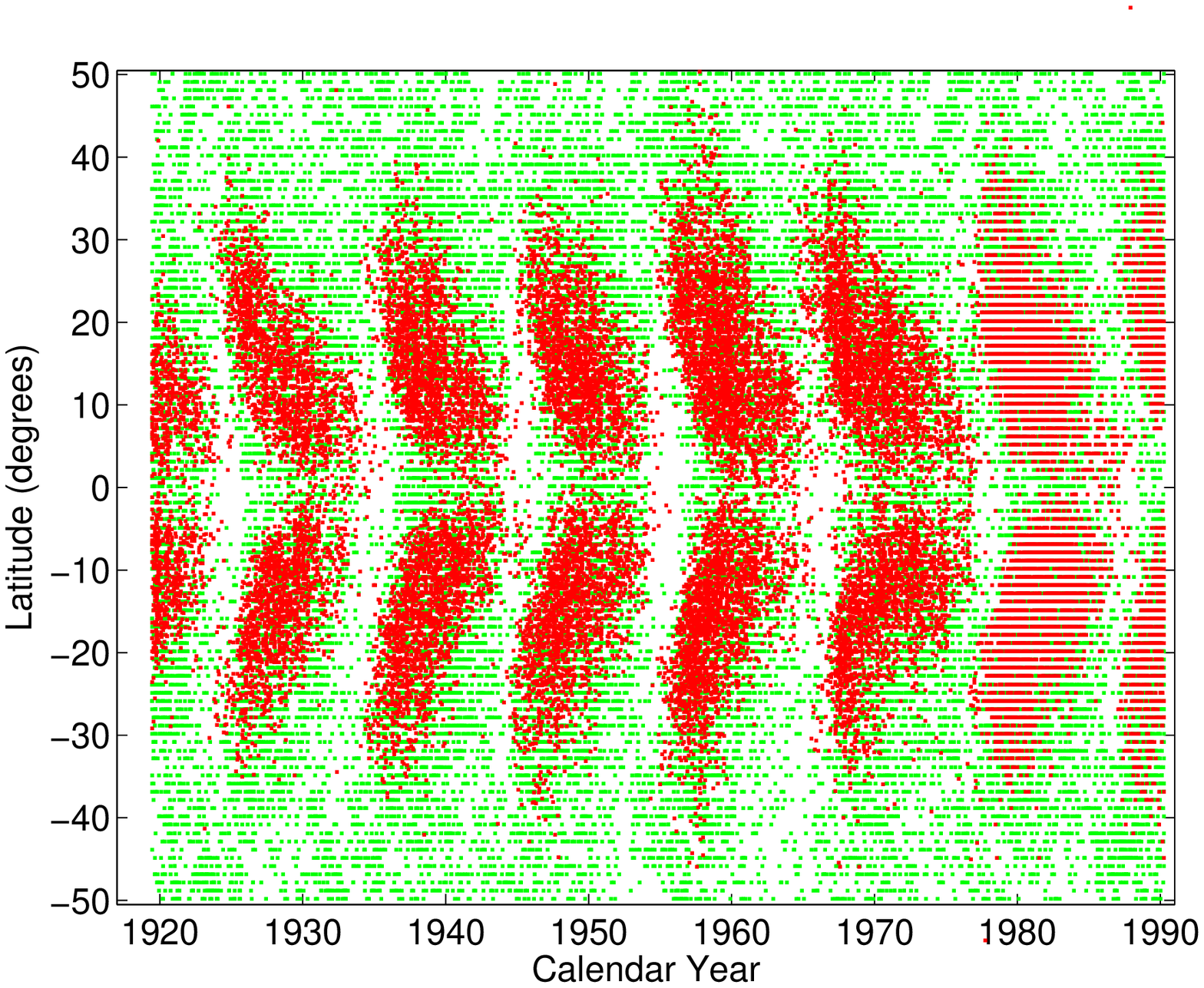} \vskip 2.5cm {\bf Figure.1 Butterfly diagrams of filaments (green dots) and sunspot
groups (red dots) from March 1919 to December 1989. }
\end{center}
\end{figure*}

\newpage
\input{epsf}
\begin{figure*}
\begin{center}
\epsfysize=12.cm\epsfxsize=12.cm \hskip -5.0 cm \vskip 30.0 mm
\epsffile{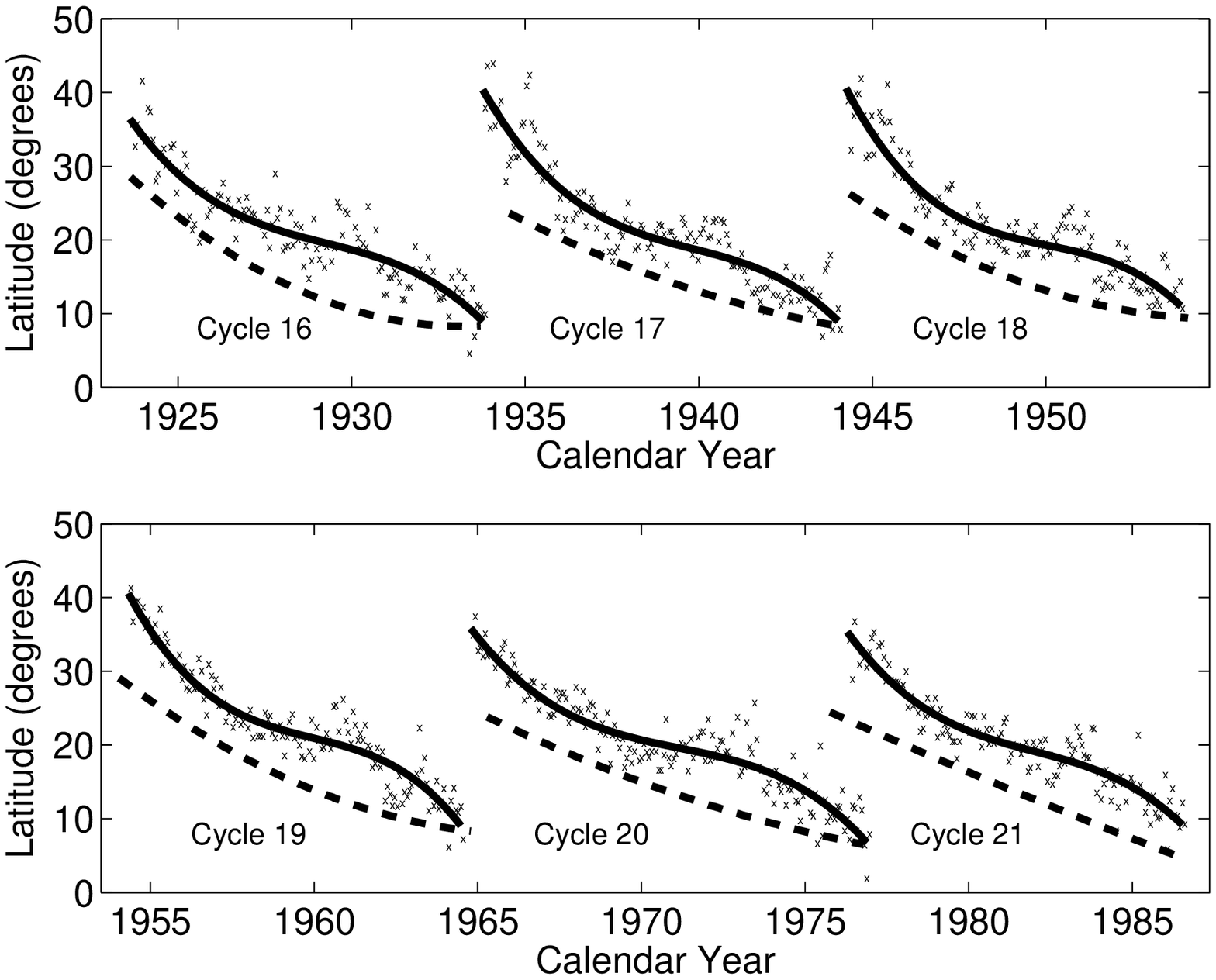} \vskip 2.5cm {\bf Figure.2 The monthly mean latitudes (crosses) of filaments in the
northern hemisphere in cycles 16 to 21 and their corresponding
third-order polynomial fits (the thick solid lines). The thick
dashed lines are the second-order polynomial fits of the monthly
mean latitudes  of sunspot groups in the northern hemisphere.}
\end{center}
\end{figure*}

\newpage
\input{epsf}
\begin{figure*}
\begin{center}
\epsfysize=12.cm\epsfxsize=12.cm \hskip -5.0 cm \vskip 30.0 mm
\epsffile{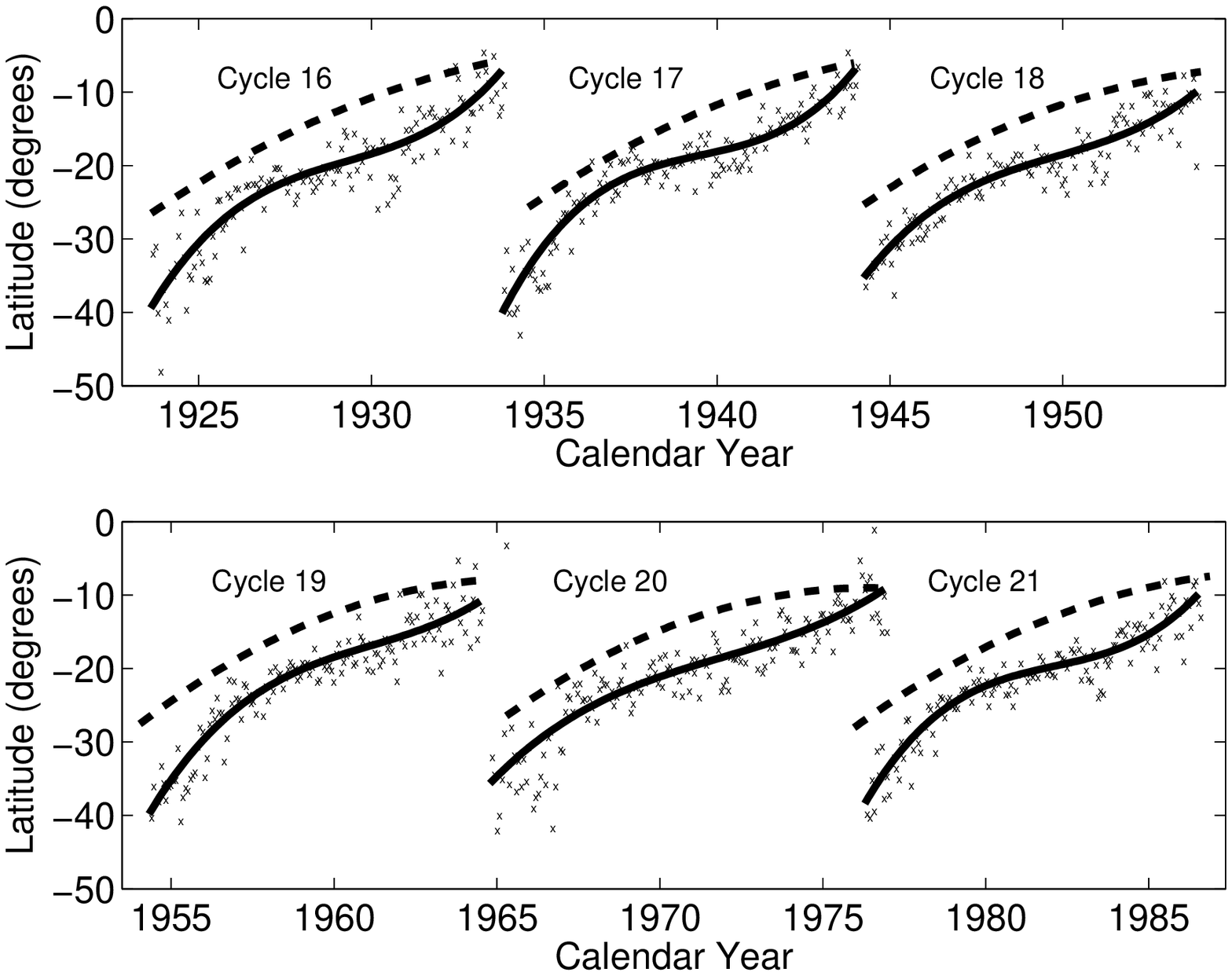} \vskip 2.5cm {\bf Figure.3 The monthly mean latitudes (crosses) of filaments in the
southern hemisphere in cycles 16 to 21 and their corresponding
third-order polynomial fits (the thick solid lines). The thick
dashed lines are the second-order polynomial fits of the monthly
mean latitudes  of sunspot groups in the southern hemisphere.}
\end{center}
\end{figure*}

\newpage
\input{epsf}
\begin{figure*}
\begin{center}
\epsfysize=12.cm\epsfxsize=12.cm \hskip -5.0 cm \vskip 30.0 mm
\epsffile{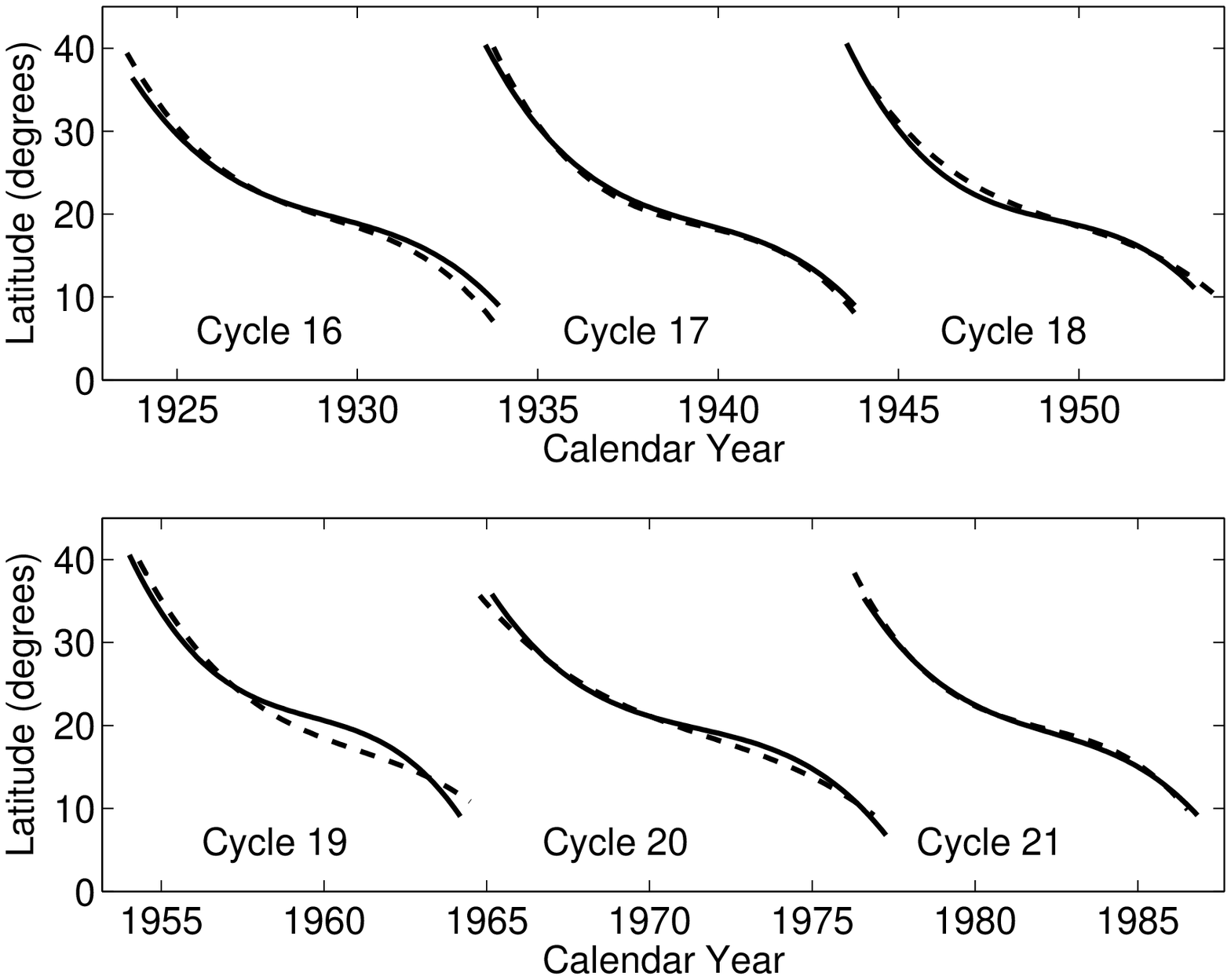} \vskip 2.5cm {\bf Figure.4 The unsigned monthly mean latitude of filaments
respectively in the northern (the solid lines)  and southern (the
dashed lines) hemispheres in a cycle for cycles 16 to 21.}
\end{center}
\end{figure*}

\newpage
\input{epsf}
\begin{figure*}
\begin{center}
\epsfysize=12.cm\epsfxsize=12.cm \hskip -5.0 cm \vskip 30.0 mm
\epsffile{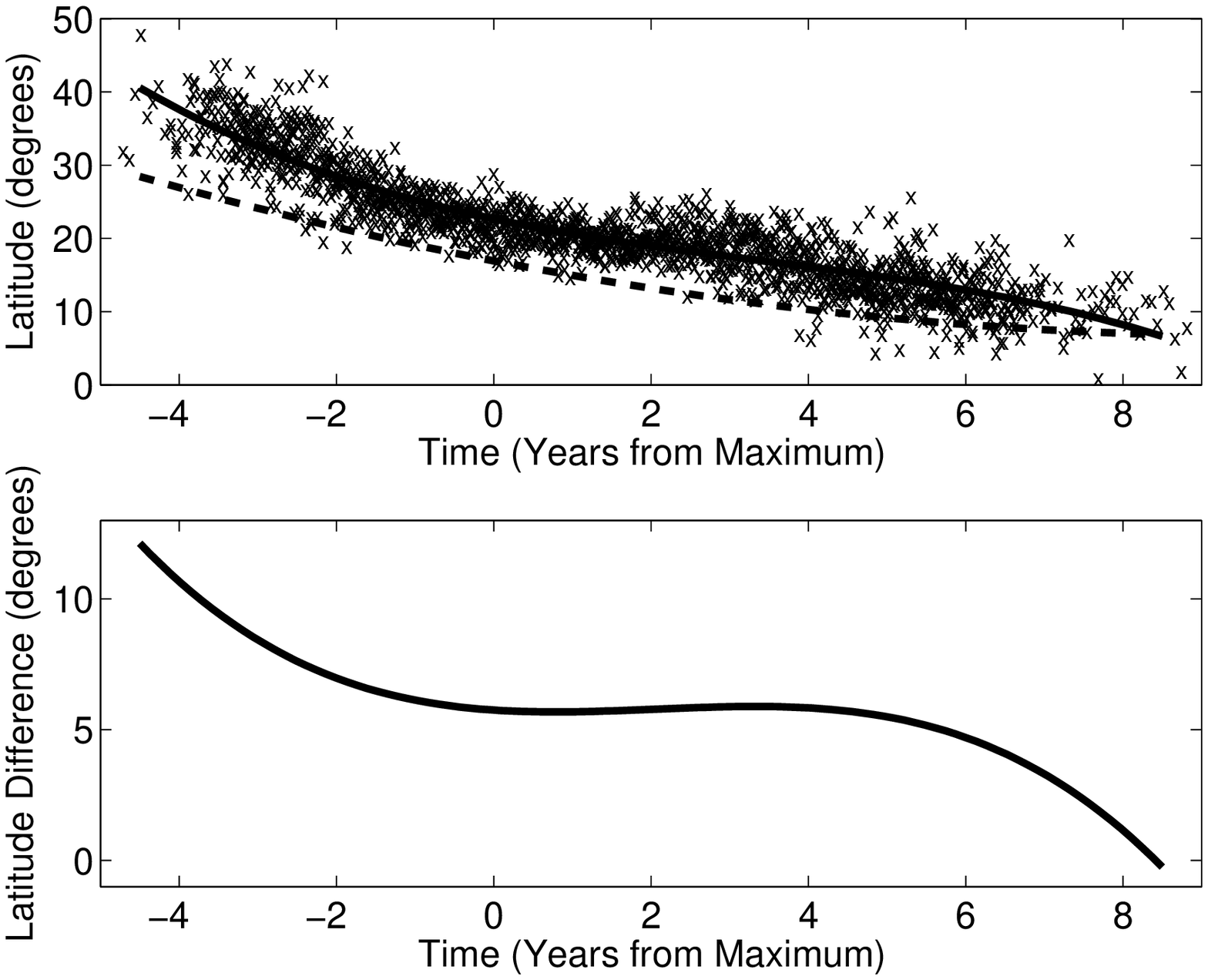} \vskip 2.5cm {\bf Figure.5 The top panel: latitude of filaments vs. time relative to
the time of sunspot cycle maximum. The unsigned monthly mean
latitude of filaments for each hemisphere and for each cycle is
plotted as an individual dot by using the time measured relative to
the maximum time of that cycle. A third-order polynomial fit to
these points is shown by the thick solid line. Similarly, the
unsigned monthly mean latitude of sunspot groups for each hemisphere
and for each cycle is plotted as an individual dot by using the time
measured relative to the  maximum  time of that cycle, and  a
second-order polynomial (the thick dashed line) is used to fit these
points. The bottom panel: the different between the thick solid and
dashed lines.}
\end{center}
\end{figure*}

\newpage
\input{epsf}
\begin{figure*}
\begin{center}
\epsfysize=12.cm\epsfxsize=12.cm \hskip -5.0 cm \vskip 30.0 mm
\epsffile{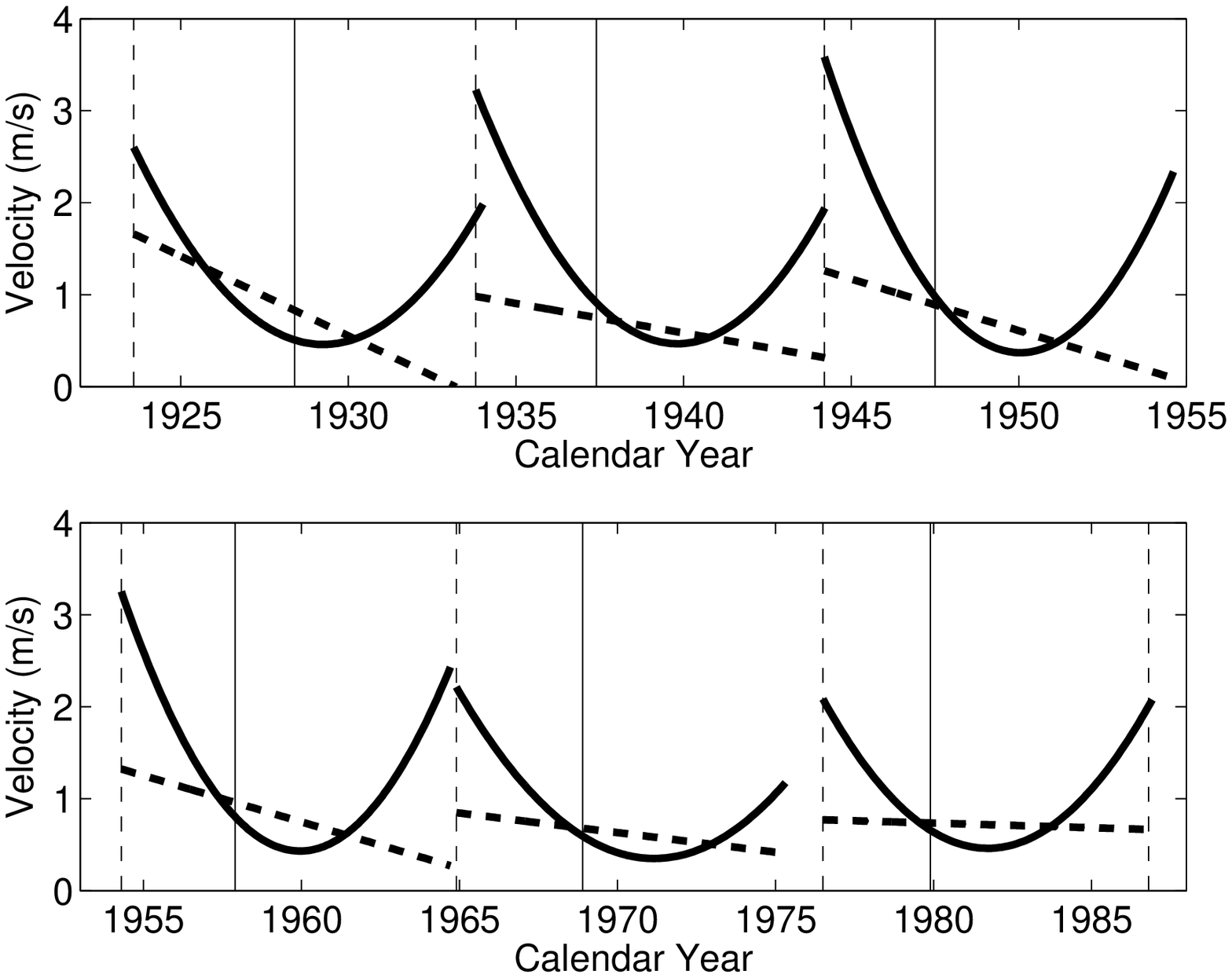} \vskip 2.5cm {\bf Figure.6 The latitudinal drift velocity respectively for filaments
(the thick solid lines) and sunspot groups (the thick dashed lines)
varying with time in the northern hemisphere in each of cycles 16 to
21. The vertical thin dashed and solid lines mark the minimum and
maximum times, respectively.}
\end{center}
\end{figure*}

\newpage
\input{epsf}
\begin{figure*}
\begin{center}
\epsfysize=12.cm\epsfxsize=12.cm \hskip -5.0 cm \vskip 30.0 mm
\epsffile{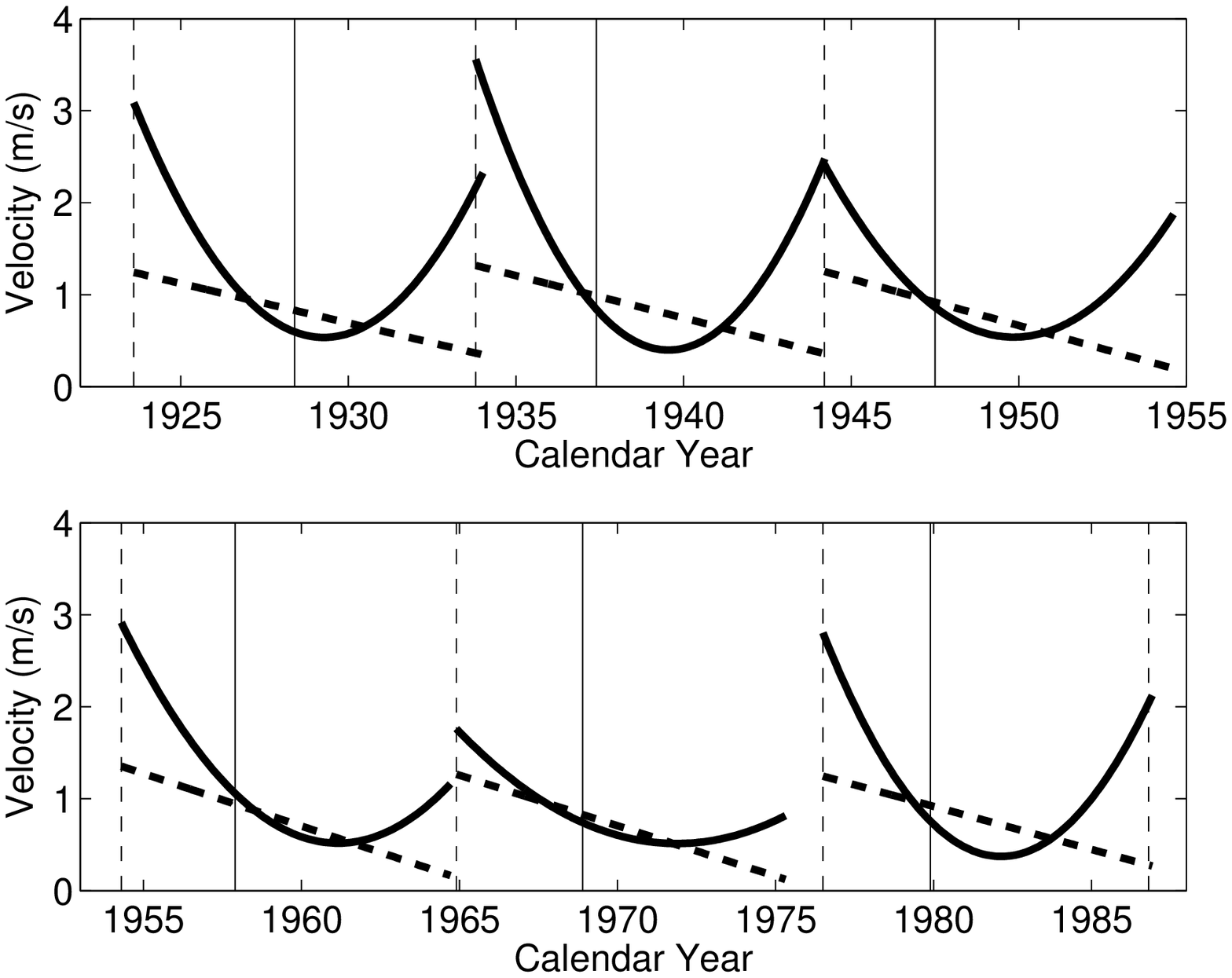} \vskip 2.5cm {\bf Figure.7 The latitudinal drift velocity respectively for filaments
(the thick solid lines) and sunspot groups (the thick dashed lines)
varying with time in the southern hemisphere in each of cycles 16 to
21. The vertical thin dashed and solid lines mark the minimum and
maximum times, respectively.}
\end{center}
\end{figure*}

\newpage
\input{epsf}
\begin{figure*}
\begin{center}
\epsfysize=12.cm\epsfxsize=12.cm \hskip -5.0 cm \vskip 30.0 mm
\epsffile{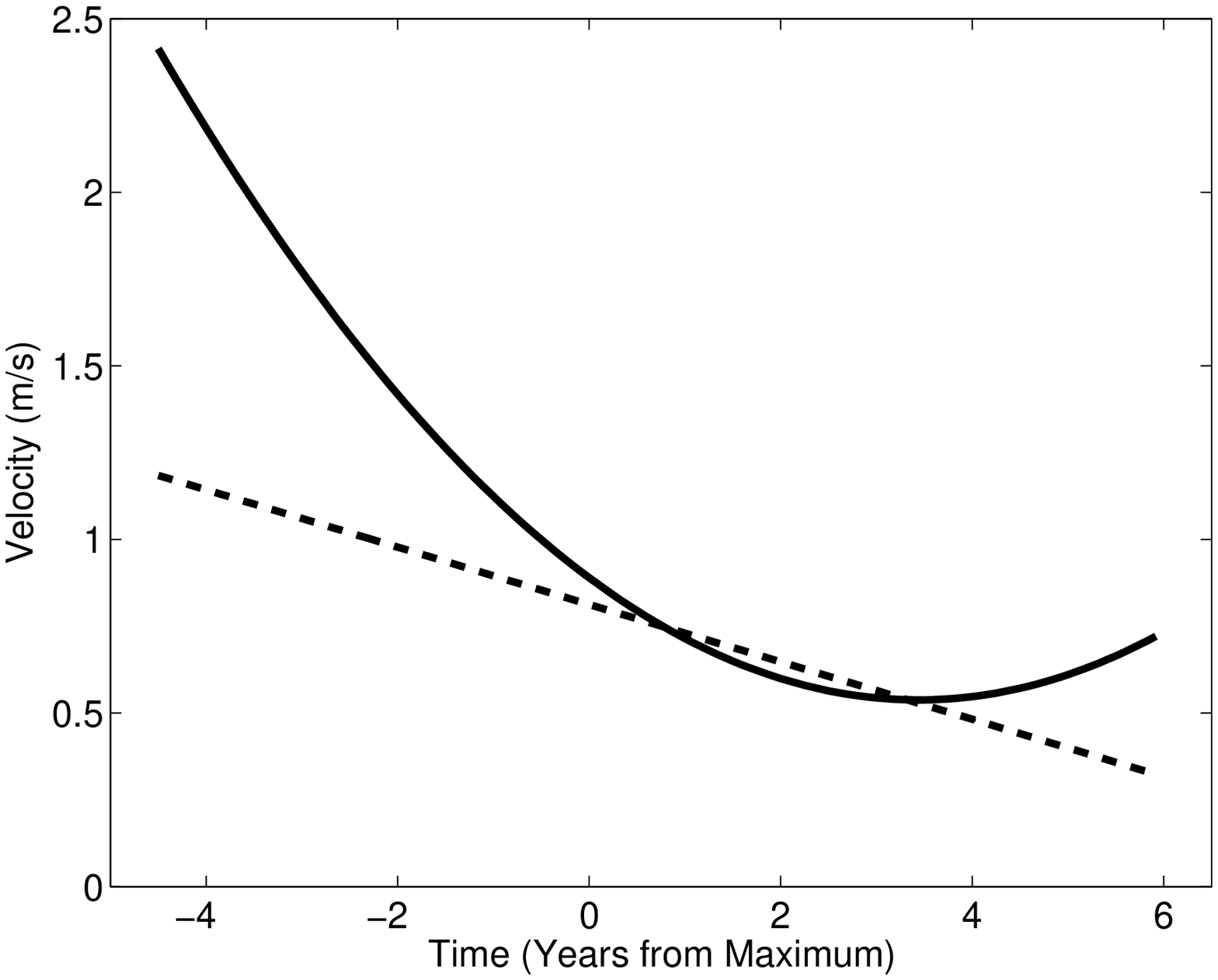} \vskip 2.5cm {\bf Figure.8 The general feature of the latitudinal drift velocity in a
cycle respectively for filaments (the thick solid line) and sunspot
groups (the thick dashed line) when the unsigned monthly mean
latitude of filaments and sunspot groups for each hemisphere and for
each cycle is plotted together by using the time
measured relative to the maximum time of that cycle (see Figure
5).}
\end{center}
\end{figure*}

\newpage
\input{epsf}
\begin{figure*}
\begin{center}
\epsfysize=12.cm\epsfxsize=12.cm \hskip -5.0 cm \vskip 30.0 mm
\epsffile{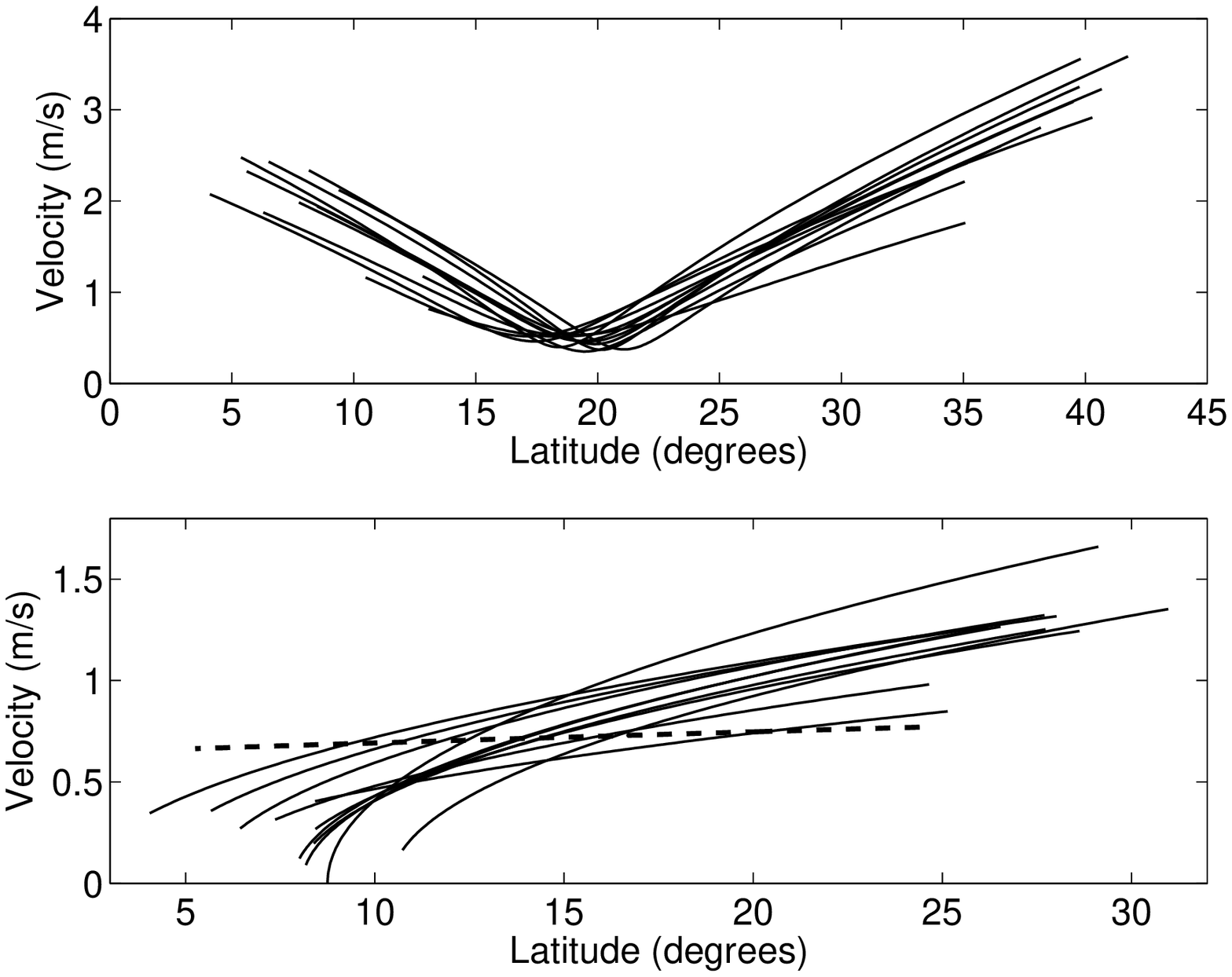} \vskip 2.5cm {\bf Figure.9 Drift velocity varying with latitude respectively for
filaments (the top panel) and sunspot groups (the bottom panel) in
each hemisphere in each cycle. The dashed lines is for sunspot
groups in the northern hemisphere in cycle 21.}
\end{center}
\end{figure*}

\newpage
\input{epsf}
\begin{figure*}
\begin{center}
\epsfysize=12.cm\epsfxsize=12.cm \hskip -5.0 cm \vskip 30.0 mm
\epsffile{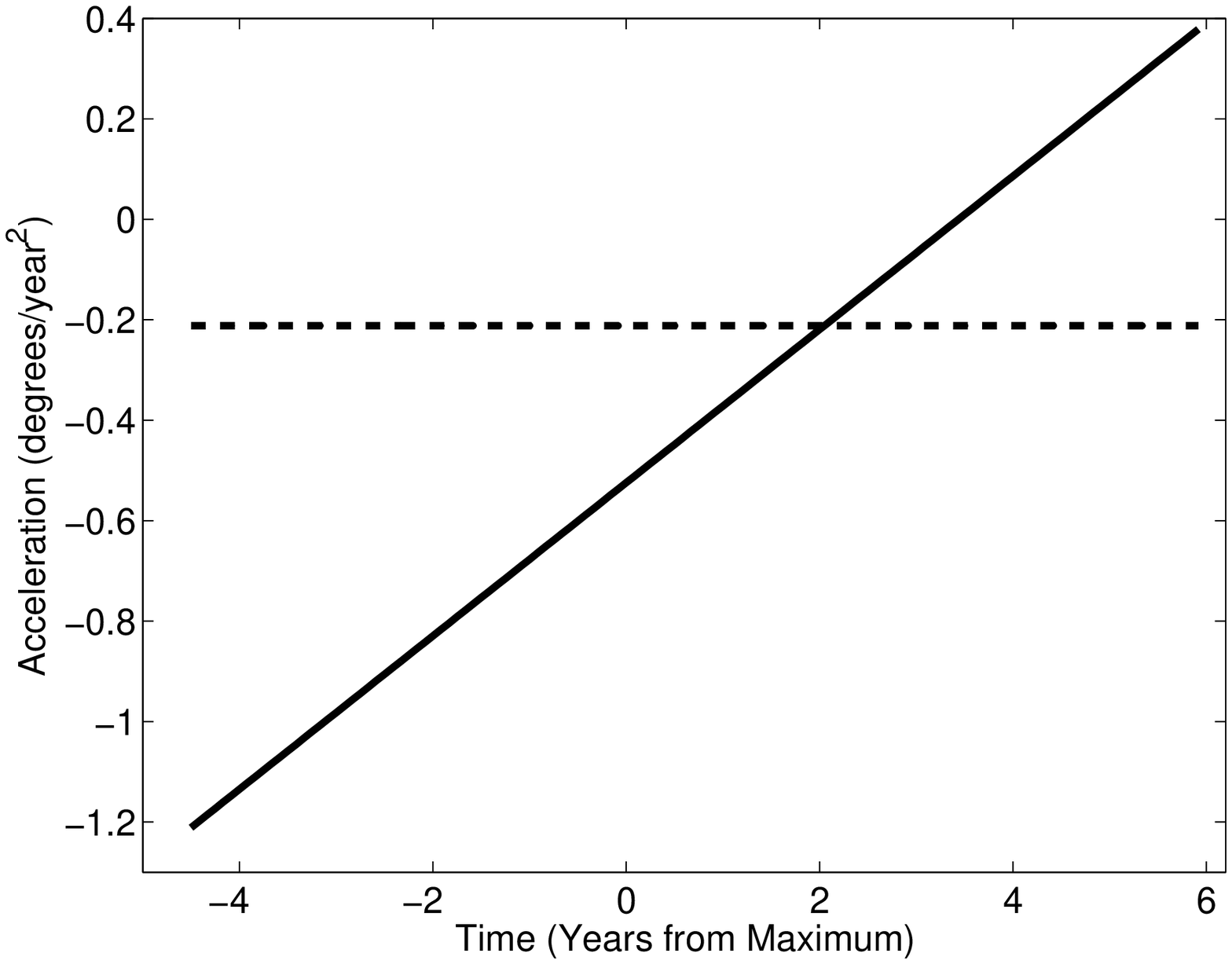} \vskip 2.5cm {\bf Figure.10 Latitudinal drift acceleration varying with time
respectively for  filaments (the thick solid line) and sunspot
groups (the thick dashed line) when the unsigned monthly mean
latitude of filaments and sunspot groups for each hemisphere and for
each cycle is plotted together by using the time
measured relative to the maximum time of that cycle (see Figure  5).}
\end{center}
\end{figure*}

\bsp \label{lastpage}
\end{document}